\documentclass[12pt,a4paper]{article}
\usepackage{amssymb}
\usepackage{appendix}
\usepackage{graphicx}

\newtheorem{theorem}{Theorem}
\newtheorem{acknowledgement}[theorem]{Acknowledgement}

\input{tcilatex}
\begin{document}

\title{The role of convective structures in the poloidal and toroidal
rotation in tokamak}
\author{Florin Spineanu and Madalina Vlad \\
National Institute of Laser, Plasma and Radiation Physics\\
Bucharest 077125, Romania}
\maketitle

\begin{abstract}
The connection between the poloidal and the toroidal rotation of plasma in
tokamak is important for the high confinement regimes, in particular in
reactor regime. The sudden onset of closed convection structures in the
poloidal section, due to the baroclinic production of vorticity, will
sustain a fast increase of the poloidal velocity and a substantial effect on
the toroidal rotation. However this is limited to the short time of the
onset transition. In real plasma however there is random generation and
suppression of convection cells and the sequence of these transient events
can prove able to sustain the effect on the toroidal rotation. We formulate
a simplified model which consists of a laminar sheared, regular, flow
situated at the boundary of a region of drift-wave turbulence. Vortical
structures, randomly generated in this turbulent region are spontaneously
advected toward the flow and are absorbed there, sustaining with their
vortical content, the sheared flow. We examine this dynamics in the
wavenumber $\mathbf{k}$ space, using reasonable approximations. We derive a
system of equations (which is a class of Devay-Stewartson system) and find
that indeed, the vortices advected and absorbed into the layer can preserve
its regular, poloidal, flow.
\end{abstract}

\section{Introduction}

Mixed regimes consisting of large scale flows (H-mode, Internal Transport
Barriers) and turbulence are expected to be the current state in ITER. The
rotation, either spontaneous or induced, will play a major role in the
quality of the confinment. The efficiency of the sheared poloidal rotation
to control the instabilities is considerably higher than that of the
toroidal rotation but it is usually assumed that the poloidal rotation
should be at the neoclassical level due to the damping induced by magnetic
pumping. This is true if the drive of the poloidal rotation (besides the
neoclassical one) relies on the Reynolds stress produced by a poloidally
-symmetric turbulence. Much higher drive of the poloidal rotation is however
provided by flows associated with the convective structures that can be
generated in the plasma cross-section beyond a threshold in the plasma
pressure gradient. This drive overcomes the damping due to the magnetic
pumping and the poloidal rotation is sustained. Cells of convection
consisting of closed, large scale flows with unique sense of rotation can be
spontaneously generated, triggered by streamers sustained by the baroclinic
term able to generate vorticity. Similar to the Reyleigh-Benard first
bifurcation (from purely conductive to convective regime), the onset is very
fast and the drive exerted on the poloidal rotation leads to a fast time
variation of the polarization radial electric field. This is sufficient to
create a distinction between the phases (first half, second half) of bounce
on a banana of trapped ions and, implicitely, leads to acceleration in the
toroidal direction. This phenomenological model has been proposed recently 
\cite{flmadi1} and the estimations are compatible with the observed effect
of reversal of the toroidal rotation in tokamak. There are several aspects
that still must be investigated and in particular the nature of
sudden-impulse of the generation of the poloidal flow as envelope of the
peripheral velocity of the convective structures. It is only on a very short
time interval (onset of convection) that the time variation of the radial
electric field can produce a toroidal effect on the bananas. If the
convection cells become saturated and stationary then the efficiency of
sustaining toroidal acceleration of bananas decreases rapidly to zero. The
process must be repeated if it is to sustain its effect on the toroidal
rotation.

On the other hand, it is highly improbable that the structure consisting of
cells of convection breaking the azimuthal symmetry can exist as a
stationary state. This is because the convection in the cells is much more
efficient as radial transport of energy than any diffusive process sustained
by small scale turbulence and the equilibrium profiles would be modified,
suppressing the origin of convection. It is more realistic to regard the
generation of convective structures as a stochastic process consisting of a
random sequence of transient events, taking place on a range of spatial
scales between $\rho _{s}$-scale drift vortices up to cells of convection
with dimensions comparable with the minor radius $a$. We should study the
way these transitory structures create the impulsive increase of the
poloidal velocity such as, on the average, to sustain the toroidal effect on
bananas. A marginal stability regime can be established, where the external
sources sustain the gradient of the pressure, the convection rolls are
formed transitorily and the excess of the gradients relative to a threshold
is transformed into pulses of heat and momentum in the radial direction.
This problem requires a treatment based on the stochastic nature of the
generation and destruction of the transversal convection cells, seen as a
random transitory process. Such a treatment simply does not exist, not even
for the famous Rayleigh-Benard system.

This problem is important \emph{per se}, beyond the phenomenological model
for the reversal of the toroidal rotation. This process has been identified
in experiments on a linear machine \cite{xu} and explained in the context of
residual stress able to sustain the coherent flow. The sequence consisting
of tilt, stretch and absorbtion of drift turbulent eddies was found to
contribute significantly to the Reynolds stress. Coalescence of strongly
tilted, almost poloidally oriented, eddies and formation of radially
periodic and poloidally closed ($m=0$) flow was proposed in \cite%
{flmadiphenom} as transition to rotation given by the stable solution of the
nonlinear equation for the electrostatic potential at the edge (the
Flierl-Petviashvili equation). The absorbtion of a coherent convection cell
inside a layer of sheared rotation is mentioned by Drake\emph{\ et al}. \cite%
{drake}. It has also been discussed in similar terms in Refs.\cite{shapiro1}%
, \cite{shapiro2}.

The structure of the paper is as follows. We formulate a model which is
relevant for the physical problem of sustainement of poloidal rotation by
stochastic rise and decay of vortices on different scales. We estimate the
average density of high amplitude vortices that are generated inside a
volume of plasma (in our case it is question of an area within the poloidal
cross-section since the third - toroidal - axis is considered irrelevant).
Then we remind the known mechanisms on motion of a vortex on a background of
gradient of vorticity, in order to see how random elements of vorticity of
only one sign will join the \ flow layer where they are absorbed. Finally,
we examine this dynamics in the wavenumber space $\mathbf{k}$ and derive a
model system of equations reflecting the interaction between the laminar
flow and the incoming drops of vorticity. This is actually the Davey
Stewartson system and indeed, the class of solutions called
\textquotedblleft solitoffs\textquotedblright\ show that the final state
exhibits unidirectional (poloidal) alignment of the resulting flow. This
confirms that the elements of vorticity coming from the turbulent region
into the laminar flow do not perturb but are conforming themselves to the
geometry of the flow, \emph{i.e.} in the poloidal direction.

\section{Formulation of the model}

The effect of advection of impulse-like convection structures (streamers,
transient convection rolls of all sizes from $\sim a$ down to the small $%
\rho _{s}$ scale of high amplitude robust drift vortices) to the layer of
sheared laminar flow and their role in sustaining the laminar flow is a
process that has aquired recognition for its importance in creating a
coherent flow pattern. In the physics of the atmosphere and in fluid physics
this is a well known phenomenon. It has been examined within the model of
potential vorticity mixing \cite{DritschellScott} for the generation of an
atmospheric jet, the equivalent of sheared laminar flow. Recent numerical
simulation indicate the interesting and unexpected random switching of the
flow between the two geometries: laminar flow and respectively a periodic
chain of positive and negative vortices \cite{BouchetSimonnet} which
indicates that these two states are indeed neighbor in the functional space
of the system's configurations. The examination of the problem in real space
is particularly difficult since it requires to analytically describe
unequivocally the emergence of a coherent structure out of turbulence. One
of the possiblity would be to derive a time-evolution equation for the
effective number of degrees of freedom that are involved in the motion, as
it results from a Karhunen - Loewe analysis. Such an approach is under
examination but with the reserve that it may not really become a
quantitative instrument.

Alternatively we can examine the system in the spectral $\mathbf{k}$ space.
The wave kinetic equation is the standard instrument to describe the
dynamics of the energy density on elementary spectral intervals (seen as 
\emph{pseudoparticles}), induced by nonlinear processes of wave turbulence 
\cite{ZakharovLvovFalkovich}. We must adapt this approach to our particular
problem, inspired by similar adaptations that have been made in several
cases in the past. The most common models for dissipative equations of
motion are the \emph{time-dependent Landau - Ginzburg models}. The model
function is $\psi $, the order parameter of the superfluid. In the absence of%
\emph{\ dissipation} the general form of the equation is \cite%
{HohenbergHalperin}%
\[
\frac{\partial \psi }{\partial t}=2i\Gamma _{0}\frac{\delta H}{\delta \psi
^{\ast }} 
\]%
where $\Gamma _{0}$ is a real constant and the hamiltonian is%
\[
H\left[ \psi \right] \equiv \int d^{d}x\left[ \frac{1}{2}\left\vert \mathbf{%
\nabla }\psi \right\vert ^{2}+\frac{1}{2}\alpha \left\vert \psi \right\vert
^{2}+\beta \left\vert \psi \right\vert ^{4}\right] 
\]

The total number of particles $N=\int d^{d}x\left\vert \psi \right\vert ^{2}$
must be constant, as well as the energy $E=H\left[ \psi \right] =$const. The
time dependent LG, is universal and occurs whenever there is a two-phase
system and a possible phase transition between them, with a function $\psi $
representing the \emph{order parameter}. Of course, such a situation is
subiacent to our more general and more complex problem. We take this as a
suggestion to \emph{simplify }our problem and reduce our objectives with the
promise that they would become more accessible: we will try to prove that
the advection of the transient convection structures to the laminar flow
layer will not perturb the geometry of the flow, which remains mainly
poloidally oriented. Then the \emph{conservation of the vorticity} will
guarantee that the content of vorticity of the individual structure joining
the laminar sheared flow will be absorbed into the new, coalesced, flow and
further will be redistributed to become homogeneous along the direction of
flow. Or, the vorticity of a laminar flow is simply the derivative of the
unidirectional velocity with respect to the transverse coordinate. The
increase of the content of vorticity in the region of the laminar flow
simply means that the total momentum (in case of a slab geometry) or angular
momentum (if the flow is confined in a poloidal layer) is increasing. We
then have a drive that may sustain the flow against the magnetic pumping.
All this now requires to prove that absorbtion of the transient convection
structures preserves the laminar flow and merge their vorticity content into
the sheared flow.

In connection with our problem of stochastic stationarity of sustainment of
poloidal flow by intermittent rise and decay of convection cells we
construct a simple model to be investigated. We consider a layer of laminar
sheared rotation where we assume that the instabilities are suppressed (this
is an ideal representation of the zonal flow or of the $H$-mode layer). The
layer is connex to a region where drift-type turbulence exists. The width of
the layer is much smaller than the region of turbulence. In the
drift-turbulent region there is stochastic generation of convective motions
and of strong vortical structures. The physical effect is different in the
two limits: for large convective events there is an effect which we see as a
jet-like impulse acting on the laminar layer; for small and robust vortices
acceding the sheared flow, the effect consists of a forcing applied to the
layers of the flow that have the same direction of flow as the rotation in
the vortex while the layers where the two directions are opposite are
decelerated. This enhances the local gradient of the velocity (\emph{i.e.}
the shear). Since the decelerated layers are in contact with the bulk,
turbulent, plasma, the angular momentum is transferred by viscosity. While
the thin laminar layer is accelerated, the large mass of turbulent plasma
will only get a low rate of compensating rotation. This ensures the
conservation of angular momentum. Although large and small vortices interact
in different ways with the sheared flow layer, we will attempt a simplified
treatment that retains the common aspects.

The shear rate (first derivative of the velocity of flow with respect to the
coordinate transversal to the layer) is assumed non-uniform, which is
equivalent to a gradient of vorticity (involving the second order
derivative). The vorticity starts from zero at the boundary with the
turbulent region and increases with the distance from it. Since the
background has a gradient of vorticity a vortex will move in a direction
which depends on the sign of its vorticity relative to the one of the
background: the \emph{clumps} (positive sign circulation vortices) are
ascending the gradient of vorticity, toward the maximum of vorticity in the
rotation layer and the \emph{holes} (negative sign circulation vortices) are
moving toward the minimum of vorticity, equivalently - they are repelled
from the sheared layer \cite{schecterdubin}. In short, vortices of only one
sign are joining the layer. The \emph{clump} vortices will be absorbed by
the sheared velocity layer with the transformation of their vorticity
content into a local enhancement of the shear of the flow. This is a source
of momentum which sustains the sheared flow, in particular sustains the
sheared poloidal rotation against the damping due to the magnetic pumping.
However it may appear as a spotaneous separation of ordered motion out of
turbulence, and would pose a problem of entropy decrease. We have to
remember that the vortices and the directed motions belonging to convection
events are actually supported, via the baroclinic terms, by the equilibrium
gradients (which are externally sustained). Similar physical processes are
probably behind the formation of the Internal Transport Barriers and the
formation of the $H$-mode rotation layer. A last comment is in order, in
connection with what appears to be a spontaneous separation of the
vorticities of the opposite signs. This is indeed what happens in the $2D$
ideal fluid at relaxation as result of the inverse cascade, and is shown by
experiments and numerical simulation \cite{montgomerysinh} and also derived
analytically \cite{flmadisinh}. Similar separation of the two signs of
vorticity has been seen for plasma \cite{KMcWT} and atmosphere, also
supported by analytical derivation \cite{flmadiatm}. The result is a final
state consisting of a pair of vortices of opposite signs that have collected
all positive and respectively all negative vorticities in the field (a
dipole). We now note that the dipolar state can be realized as either two
distinct vortices of opposite signs, or, in a circular geometry, as a
central part with vorticity of one sign surrounded by an annular region of
vorticity of the opposite sign. Transitions are possible between the two
"dipolar" geometries of vorticity separation and the change is very fast. We
only mention this facts in support of our phenomenological representation of
the sustainment of the sheared rotation, but we will not develop this
subject here.

We take the layer of plasma flow in the $y$ ($\equiv $\ longitudinal,
streamwise, poloidal in tokamak) direction with sheared velocity varying in
the $x$ ($\equiv $\ transversal on the layer of flow, radial in tokamak)
direction. It is described by a scalar function $\widehat{A}\left(
x,y,t\right) $ which is the complex amplitude of slow spatial variation of
the envelope of the eddies of drift turbulence. From $\widehat{A}\left(
x,y,t\right) $ the components of the velocity are derived as $\mathbf{v}%
_{E}=-\mathbf{\nabla }\widehat{A}\left( x,y\right) \times \widehat{\mathbf{n}%
}/B$ where $\widehat{\mathbf{n}}$ is the versor of the magnetic field. We
also define a scalar function $\phi \left( x,y,t\right) $ which represents
the field associated to the incoming vortices that will interact and will be
absorbed into the sheared, laminar, flow, transfering to it their vorticity
content. The process in which a localized vortex is peeled-off and absorbed
into the sheared flow may be seen as a transfer of energy in the spectrum
from the small spatial scales involved in the structure of the vortex toward
larger spatial scales of the sheared flow, with much weaker variations on
the poloidal direction. This can be seen as a propagation in $\mathbf{k}$
space.

We must first estimate what is the number of high amplitude, robust vortices
that are generated by the drift-turbulence in a unit area. This problem is
treated in the Appendix 1 where it is shown that the density of strong
vortices is $n=3/\left( \pi \lambda ^{2}\right) $ where $\lambda $ is the
typical size of the drift eddies normalized to $\rho _{s}$.

\section{How these vortices ascend to the flow layer?}

The answer is: due to the existence of the background gradient of vorticity.
This has been clearly described for non-neutral plasma in \cite%
{schecterdubin} and is also known in the physics of fluids \cite{marcus},
the physics of the atmosphere \cite{wang} and in astrophysics (accretion
disks) \cite{linbarrancomarcus}.

Since the background has a gradient of vorticity a vortex will move in a
direction which depends on the sign of its vorticity relative to the one of
the background:

\begin{itemize}
\item the prograde vortices are moving toward the maximum of vorticity and

\item the retrograde vortices are moving toward the minimum of vorticity.
\end{itemize}

The prograde vortices eventually are absorbed by the sheared layer and they
contribute with their vorticity content to the momentum of the background
flow. This is a source of momentum which sustains the sheared flow, in
particular sustains the sheared poloidal rotation against the damping due to
the magnetic pumping.

A \emph{clump} of vorticity (positive circulation $\Gamma _{v}>0$) goes up
the gradient of background vorticity. For a sheared velocity flow with
poloidal velocity $v_{\theta }$ increasing linearly with radius, with a
coefficient $Q$%
\[
v_{\theta }=Qr 
\]%
but in a layer which has background vorticity $\zeta _{0}^{\prime }\ \ $along%
$\ \ r$ , the radial motion of the positive vortex (position $r_{v}$) is%
\[
\frac{dr_{v}}{dt}=\zeta _{0}^{\prime }l^{2}\ln \left( \frac{cr_{v}}{l}%
\right) \arctan \left( \left\vert Q\right\vert t\right) 
\]%
where $l\equiv \sqrt{\left\vert \Gamma _{v}/\left( 2\pi Q\right) \right\vert 
}$ \cite{schecterdubin2}. Schecter and Dubin \cite{schecterdubin} find $%
c\lesssim 1$. The position of the vortex is greater but comparable to the
typical length $l$, $r_{v}/l\gtrsim 1$. The $\arctan $ function is bounded
by $\pi /2$.

The negative-circulation vortices, $\left( \Gamma _{v}<0\right) $ called 
\emph{holes} are moving in the opposite direction therefore they remain in
the turbulent region and are repelled by the layer of sheared flow with
global conservation of angular momentum. This means that only vortices of a
certain sign ascend to the flow and are absorbed in it. In the following we
will analyse this process as an interaction between the two fields: $%
\widehat{A}\left( x,y,t\right) $ and the field representing the incoming
vortices.

\section{Dynamics in the $\mathbf{k}$ - space of the interaction between the
laminar flow field and the field of the vortical cells}

In order to derive a dynamic equation for the scalar streamfunction of the
sheared flow $\widehat{A}\left( x,y,t\right) $ we remind the properties
resulting from multiple space-time analysis of a turbulent field, like the
drift waves interacting with a coherent, sheared flow. The typical equation
in the hydrodynamic description is Rayleigh-Kuo or \emph{barotropic}
equation and the multiple space-time scale analysis leads to the Nonlinear
Schrodinger Equation for the amplitude $W\left( x,y,t\right) $ of the \emph{%
envelope} of the oscillatory waves excited in the fluid \cite{flmadi2}. The
general structure of this equation is%
\begin{equation}
i\frac{\partial W\left( x,y,t\right) }{\partial t}=\Delta W\left(
x,y,t\right) -\gamma \left\vert W\right\vert ^{2}W^{\ast }  \label{eq1}
\end{equation}%
The interesting term is the nonlinearity (the last term) which may be seen
as arising from a self-interaction potential $V\left[ W\left( x,y,t\right) %
\right] $ (see the next Section). The later potential results naturally from
the multiple scale analysis and it is the effect on the envelope amplitude
of the nonlinearity consisting of convection of the vorticity by its own
velocity field. We must work with an amplitude $A\left( k_{x},k_{y},t\right) 
$ in the spectral $\mathbf{k}-$space, which we see as the envelope of
oscillations representing propagation of perturbation in the spectrum, this
propagation being a reflection of what is the process of the incoming vortex
dissolving itself and being absorbed by the flow, - in the physical space.
Looking for an analytical model for $A$ in the $\mathbf{k}$-space we will
use the structure which is suggested above.

We expect to have a diffusion in $\mathbf{k}-$space and the same
self-limitting nonlinear term%
\begin{equation}
i\frac{\partial A\left( k_{x},k_{y},t\right) }{\partial t}=\left( \frac{%
\partial ^{2}}{\partial k_{x}^{2}}+\frac{\partial ^{2}}{\partial k_{y}^{2}}%
\right) A\left( k_{x},k_{y},t\right) -\gamma \left\vert A\right\vert
^{2}A^{\ast }+S  \label{eq2}
\end{equation}%
and we have added a source $S$ that represents the interaction of the basic
flow $A$ with the vortices that are randomly generated in the drift
turbulence and are joining the flow.

The addition of \textquotedblleft drops\textquotedblright\ of vorticity is
represented by the $\mathbf{k}$-space scalar function $\phi $%
\begin{equation}
\phi \left( k_{x},k_{y},t\right)  \label{eq3}
\end{equation}%
Several space scales are present in the structure of the function $\phi
\left( x,y,t\right) $ in real space. The reason is that we must consider
that the vortices that are continuously generated in the turbulent region
have two characteristics:

\begin{itemize}
\item the space extension of individual vortices coming into the layer has a
wide range of values. The range extends from vortices of the dimension $\rho
_{s}$ (the natural result of excitations sustained by the ion-polarization
drift nonlinearity) up to convective events representing streamers with
-closed (roll-type) geometry and spatial extension given by the inverse of $%
k_{\theta }=m/r$ for $m$ of only few units.

\item the spatial distribution of the place where these vortical structures
are reaching the layer of sheared flow is random over practically all the
circumference of the poloidal flow.
\end{itemize}

We will draw conclusions about the $\mathbf{k}$-space profile of $\phi $
based on the contributions coming in from various scales in the real space.
The most simple description neglects the random aspect of the content of $%
\phi $ and simply ennumerates the possible spatial scales. For this we adopt
a discrete representation of the dimensions of vortices, starting from a
minimal size of the order of $\rho _{s}$ up to a low-$m$ fraction of the
circumference $2\pi r$ of the poloidal rotation layer, this last scale
corresponding to large convection roll-type events.

The above elementary assumption, \emph{i.e.} the discrete number of space
extension of vortices, implies that the function $\phi \left(
k_{x},k_{y}\right) $ presents higher amplitudes around wavenumbers $\mathbf{k%
}$ which correspond to the inverses of the spatial scales. The simplifying
assumption which takes uniform spatial distribution of the events of arrival
of vortices (of any scale) to the layer implies periodicities on various
spatial scales.

We consider separately the $y$ direction (poloidal). One periodicity can be
at the level of a smallest vortex, $k_{y}^{\left( 1\right) }$. This means
that a sequence of maxima of $\phi \left( x,y\right) $ appear with spatial
periodicity $2\pi /k_{y}^{\left( 1\right) }$ and this corresponds to the
addition of smallest vortices into the flow layer. Further, on a longer
space scale there is another periodicity, $k_{y}^{\left( 2\right) }$. This
means that the sequence on the lower scale (with much denser spatial
granulations of vorticity, $2\pi /k_{y}^{\left( 1\right) }$) is modulated on
a longer space scale with a periodicity $2\pi /k_{y}^{\left( 2\right) }$.
This is because larger vortices (but not yet convection rolls) are absorbed
into the flow layer. Large vortices carry smaller vortices. These
considerations can be extended to other, intermediate, levels of
periodicity. Finally we can have the largest (accessible) spatial scale $%
2\pi /k_{y}^{\left( n\right) }$ where a convective event occurs as a result
of a stochastic event, or a streamer sustained by the baroclinic term and
having the close-up property.

We note that the periodicity in the poloidal direction $y$ cannot occur
without a corresponding periodicity in the $x\equiv $ radial direction, as
far as the vortices and convective events are concerned. It is reasonable to
assume a linear dependence : small scale periodicities in the poloidal
direction (small scale robust drift vortices) occur on a similar radial
scale, \emph{i.e.} again small periodicity in the $x\equiv $ radial
direction. With larger $y$ wavelengths, we expect to involve also larger
radial wavelengths since the physical process involves more fluid motion in
the closed-up convection cell.

This suggests a proportionality between the periodicities on the poloidal
direction, $y$, and the periodicities on the radial, $x$, direction, in the
structure of $\phi \left( k_{x},k_{y}\right) $. This means that the Fourier
components along $y$ (\emph{e.g. }at $k_{y}$) are proportional with the
Fourier components along $x$ (at $k_{x}$) with the same proportionality
coefficient on all the spectral interval. The fact that there is (assumed)
proportionality on the periodicities on $y$ and on $x$, requires for
analytical description a hyperbolic operator%
\begin{equation}
\left( \frac{\partial ^{2}}{\partial k_{x}^{2}}-\frac{1}{a^{2}}\frac{%
\partial ^{2}}{\partial k_{y}^{2}}\right) \phi \left( k_{x},k_{y}\right)
\approx 0  \label{eq4}
\end{equation}%
where the D 'Alamebrtian operator $\square _{k_{x}k_{y}}\equiv \frac{%
\partial ^{2}}{\partial k_{x}^{2}}-\frac{1}{a^{2}}\frac{\partial ^{2}}{%
\partial k_{y}^{2}}$ is the signature of the correlated periodicities on the
poloidal and radial directions of the field $\phi $ representing the
incoming drift vortices - up to convection rolls, generated in the drift
turbulence region, moving against the gradient of vorticity, and finally
absorbed into the rotation layer. The \ "velocity " $a$ is the ratio of the
periods on the two $k$-space directions and we take it $a\approx 1$. The
centers of the vortices reach points that belong to the lines $y=\pm ax$.
The general form of the solution of the homogeneous equation is%
\begin{equation}
\phi \left( k_{x},k_{y}\right) \sim \exp \left( ik_{x}x+ik_{y}y\right)
\label{eq5}
\end{equation}

When $\phi $ is \ "free" it is double periodic and it verifies the above
equation. However the destruction (\emph{peeling-off}) of the incoming
vortices and their absorbtion through interaction with the background flow, $%
A$, is the loss of the double periodicity. The departure relative to the $%
\mathbf{k}$-space double periodicity (the \ "free" state of $\phi $) feeds
the flow field $A\left( k_{x},k_{y},t\right) $ such as to enhance it in
spectral regions comparable to those where the vortices are localised, \emph{%
i.e.} higher $\left\vert \mathbf{k}\right\vert $. Basically, if the free $%
A\left( k_{x},k_{y},t\right) $ is localised on $k_{y}$ close to $k_{y}\sim 0$
(equivalent to an uniform poloidal flow), the source induced by $\phi $ is a
charge which acts through a Poisson equation on the \ "potential" given by
the amplitude $AA^{\ast }$. We should recall that we want to represent the
physical process in which incoming vortices transfer via absorbtion their
vortical content to the sheared flow, sustaining and/or enhancing the
poloidal, $y$, rotation. This is mainly manifested as the evolution of an
initial oscillation generated by $\phi $ on the $k_{y}$-spectrum with
collection of the energy in a region close to $k_{y}\sim 0$, since this
corresponds in real space to the uniformization of the flow in the poloidal
direction, after an event of absorbtion of a vortex. Then, ignoring the less
significant periodicity in $k_{x}$ direction, corresponding to periodic
perturbation propagating in the $x$ (radial) direction, the interaction may
be represented as%
\begin{equation}
-\beta \frac{\partial ^{2}}{\partial k_{y}^{2}}\left\vert A\right\vert
^{2}\approx \left( \frac{\partial ^{2}}{\partial k_{x}^{2}}-\frac{\partial
^{2}}{\partial k_{y}^{2}}\right) \phi \left( k_{x},k_{y}\right)  \label{eq6}
\end{equation}

The constant $\beta $ is a measure of the permitivity of the equivalent
electrostatic problem. Using the specific terminology $\beta $ measures the
decrease of the effectiveness of the effect of the charge (right hand side
of the Eq.(\ref{eq6})) on the potential $\left\vert A\right\vert ^{2}$ due
to the \textquotedblleft polarization\textquotedblright . Essentially, the
amount of destruction of the incoming vortex, measured by the departure from
the \ \textquotedblleft free\textquotedblright\ state by the right hand side
of the Eq.(\ref{eq6}) is transferred to the background flow by first
exciting elementary waves in real space, while $A$ is only an envelope. The
perturbation of the sheared flow $A$ (via the left hand side of the Eq.(\ref%
{eq6})) results from the nonlinear interaction of these excited waves, \emph{%
i.e.} by convecting the modified vorticity with the background velocity
flow. This is the nonlinear term of the right-hand side of Eq.(\ref{eq2}).
Then it is reasonable to assume $\beta \approx \gamma $.

Returning to the equation for $A\left( k_{x},k_{y},t\right) $ we specify the
interaction between the two fields $A$ and $\phi $ in the simplest way%
\begin{equation}
S=2\phi \left( k_{x},k_{y},t\right) A\left( k_{x},k_{y},t\right)  \label{eq7}
\end{equation}%
Then%
\begin{equation}
i\frac{\partial A\left( k_{x},k_{y},t\right) }{\partial t}=\left( \frac{%
\partial ^{2}}{\partial k_{x}^{2}}+\frac{\partial ^{2}}{\partial k_{y}^{2}}%
\right) A\left( k_{x},k_{y},t\right) -\gamma \left\vert A\right\vert A^{\ast
}+2\phi A  \label{eq8}
\end{equation}%
to which we add the equation for $\phi $,%
\begin{equation}
\left( \frac{\partial ^{2}}{\partial k_{x}^{2}}-\frac{\partial ^{2}}{%
\partial k_{y}^{2}}\right) \phi \left( k_{x},k_{y},t\right) \approx -\gamma 
\frac{\partial ^{2}}{\partial k_{y}^{2}}\left\vert A\right\vert ^{2}
\label{eq9}
\end{equation}

These two equations are known as Davey-Stewartson system and our case is
DS-I \cite{chow}. It is exactly integrable and several analytical solutions
are available, in terms of Riemann theta functions or Jacobi elliptic
functions. Of particular importance for our physical problem is the \emph{%
long wave limit} solutions, as obtained by Chow using the Hirota method. We
reproduce here his result \cite{chow}, with details given in Appendix B.%
\begin{equation}
A=\frac{r\left( 1-k\right) }{2}\left[ \frac{\tanh \left( sk_{x}\right) -%
\sqrt{k}\mathrm{sn}\left( rk_{y},k\right) }{1+\sqrt{k}\mathrm{sn}\left(
rk_{y},k\right) \tanh \left( sk_{x}\right) }\right] \exp \left( -i\Omega
t\right)  \label{eq10}
\end{equation}%
\begin{eqnarray}
\phi &=&2r^{2}\left( 1-\frac{E}{K}\right)  \label{eq11} \\
&&-2r^{2}\frac{k\left( k\mathrm{sn}^{2}\left( rk_{y},k\right) +\tanh
^{2}\left( sk_{x}\right) \right) +\left( 1+k^{2}\right) \sqrt{k}\mathrm{sn}%
\left( rk_{y},k\right) \tanh \left( sk_{x}\right) }{\left( 1+\sqrt{k}\mathrm{%
sn}\left( rk_{y},k\right) \tanh \left( sk_{x}\right) \right) ^{2}}  \nonumber
\end{eqnarray}%
These solutions are called \emph{solitoffs} and they decay in all directions
except a preferred one. They are \ "semi-infinite solitary waves". For
example, for $k=0.7$ and $r=1$ the solution shows a peak of $\left\vert
A\right\vert ^{2}$ around $k_{y}\approx 0$ which in physical terms means
quasi-uniform flow along the poloidal direction. This is compatible with
what we expect from this analytical model: the turbulence advects
\textquotedblleft drops\textquotedblright\ of vorticity into the rotation
layer and this does not destroy the flow but sustains it. It is actually a 
\emph{discrete} Reynolds stress, or a process which is reversed version of
the usual Kelvin-Helmholtz instability.

\bigskip

\section{Discussion}

The Figure (1) represents the amplitude of the solution given in Eq.(\ref%
{eq10}). We now want to understand in more detail what was the origin of the
final spectral structure (\emph{i.e.} the solution written above) in our
previous analysis, which was mostly qualitative and based on several
assumptions.

We note that the solution looks compatible with our idea about the build-up
and/or sustainment of a sheared laminar flow out of random convection of
momentum from the deep region where turbulence is active. We first refer to
the term $-\gamma \left\vert A\right\vert ^{2}A^{\ast }$. Essentially this
term is the quantitative translation of the physical fact that the flow,
after a perturbation (which later we will specify as "absorbing" an incoming
vortex), will try to re-establish the uniformity of the flow by spreading
the new perturbing vorticity as shear of the flow. Although this is a very
approximative image we expect that this term in the action functional (\emph{%
i.e.} in the integral of the Lagrangian density) to penalize departures of
the flow configurations from uniformity of the flow. The departures are
measured as the square of the difference between the actual magnitude of $%
AA^{\ast }$ and a reference uniform configuration. Then the system moves in
the function space in a potential $V\left[ \varphi \left( \mathbf{k}\right) %
\right] $ with two extrema $\pm v$ (symmetric relativ to $\varphi \left( 
\mathbf{x},t\right) \equiv 0$), so that in its Lagrangian there is a term $%
\mathcal{L}^{nl}$ of nonlinear self-interaction of the scalar field $\varphi 
$, of the general type%
\begin{eqnarray*}
\mathcal{L}^{nl} &\sim &\int dtd^{2}x\left( \varphi \varphi ^{\ast
}-v^{2}\right) ^{2} \\
&=&\int dtd^{2}x\int dt^{\prime }d^{2}x^{\prime }\left[ \varphi \left( 
\mathbf{x},t\right) \varphi ^{\ast }\left( \mathbf{x},t\right) -v^{2}\right]
\\
&&\times U\left( \mathbf{x},t;\mathbf{x}^{\prime },t^{\prime }\right) \left[
\varphi \left( \mathbf{x}^{\prime },t^{\prime }\right) \varphi ^{\ast
}\left( \mathbf{x}^{\prime },t^{\prime }\right) -v^{2}\right]
\end{eqnarray*}%
with $U\left( \mathbf{x},t;\mathbf{x}^{\prime },t^{\prime }\right) \equiv
\delta \left( \mathbf{x}-\mathbf{x}^{\prime },t-t^{\prime }\right) $. This
means that the stationary states will try to attain one of the equilbria
given by the extremum of $V\left[ \varphi \right] $. In our simplified case
the "reference configuration" that the system takes as ground state above
which it penalizes any departure is simply $v\equiv 0$. This means that we
represent the simultaneous wiping out of the local perturbation \emph{and}
the suppression of the shear, without having any reason for the latter,
except for tractability. The potential $\delta \left( \mathbf{x-x}^{\prime
},t-t^{\prime }\right) $ makes the uniformity ($v\equiv 0$ in this
simplified description) an attractive state, if no source term is present.
With $v\equiv 0$ the nonlinearity becomes the same as in the Nonlinear
Schrodinger Equation, as we have mentioned above, in the construction of the
model.

On the other hand we see that in the presence of a \emph{source} the
solution for fixed, greater than zero, $k_{x}$, is not exactly at the
"poloidal uniform flow" $k_{y}=0$ but shifted, the spectrum is peaked in $%
k_{y}$ at a small but finite value. This means that our description actually
obtains longwave poloidal oscillations but not exactly symmetric $m\equiv 0$
poloidal flow. This shift is due to the \emph{average} of the spatial
extensions of the incoming perturbations and is dependent on the scale
factor ('wavelength') $r$. The two parameters $r$ and $s$ must be seen as
representative space scales along the poloidal $\left( y\right) $
respectively radial $\left( x\right) $ directions.

We have simply neglected the structure of the spectrum on $k_{x}$ (radial).
In the solution it occurs with a uniform $k_{x}$-space contributions for
positive wavenumbers (the $\tanh $ function rises rapidly after a space
scale $s$ and remains constant after that for all $k_{x}>1/s$). This means
that the structure of the flow on the transversal direction (a cross section
along the radius) contains all possible motions and modulations, without
however placing emphasis on any of them. It becomes effective the assumption
that the convections and vortices are coming in all sizes, and we have
neglected the second derivative of the flow amplitude $A$ to $k_{x}$ in Eq.(%
\ref{eq6}), compatible with the approximate independence of $A$ on $k_{x}$.

We note that the identical reproduction of the localized maxium close to $%
k_{y}\approx 0$ at periodic intervals along $k_{y}$ axis is the result of
the periodicity%
\[
sn\left( rk_{y},k\right) =sn\left( rk_{y}+4K,k\right) 
\]%
of the Jacobi elliptic function $sn$, where 
\begin{eqnarray*}
k &=&2s/r-1\ \ \text{and} \\
K\left( k\right) &=&\int_{0}^{\pi /2}\frac{d\xi }{\sqrt{1-k^{2}\sin ^{2}\xi }%
}
\end{eqnarray*}
For slightly elongated perturbations along the direction of the flow
(poloidal) 
\[
s\lesssim r 
\]%
we have%
\[
k\lesssim 1\ \ \rightarrow \ \ K\rightarrow \infty 
\]%
which means that the periodicity does not appear to be relevant in our case.

\section{Conclusion}

In conclusion, this simple analytical analysis supports the idea that the
randomly generated vortical motions, including the convection rolls that
rise and decay transiently in the plane transversal to the magnetic field,
are able to sustain poloidal rotation. When the time scale is very short, as
for convection rolls driven by the \emph{baroclinic} effect, the poloidal
flow arising as envelope of the rolls has fast time variation and the
neoclassical polarization effect acts effectively on the toroidal rotation.
This process accompanies every event consisting of configuration of a closed
convection pattern. Then, even if these events are random, their statistical
average is an effective source of poloidal flow.

\bigskip

\begin{acknowledgement}
Work supported partially by the Contracts BS-2 and BS-14 of the Association
EURATOM - MEdC Romania. The views presented here do not necessarly
represents those of the European Commission.
\end{acknowledgement}

\bigskip

\bigskip 
\begin{appendices}
\section{Appendix. Estimation of the density of high amplitude vortices generated in the drift-wave turbulence}
\renewcommand{\theequation}{A.\arabic{equation}} \setcounter{equation}{0}

In this Appendix we discuss the problem of estimating the average number of
strong vortices that are generated in a two-dimensional region of
drift-turbulent plasma. Being a complex problem, it will be the subject of a
separate work. We limit ourselves here to mention the main steps of this
calculation, revealing the necessity to link apparently different systems.

We adopt the definition that a strong, robust vortex of high amplitude is
one which is a physical realisation of the mathematical singular vortex, the
latter being characterised by $\omega \left( \mathbf{r}\right) \sim \delta
\left( \mathbf{r}\right) $. This approximation is useful since we will now
look for the statistically averaged density of singular vortices in a region
of turbulence.

The following connections can be made:

\begin{enumerate}
\item The drift wave turbulence in the hydrodynamic regime and at spatial
scales that are much larger than $\rho _{s}$ is governed by the
cuasi-three-dimensional Hasegawa-Mima equation. However, for strong magnetic
field, there is not too much difference to ignore the intrinsic length, the
Larmor radius, and to consider the approximate description given by the
scale-free hydrodynamic model, \emph{i.e.} the $2D$ Euler equation%
\[
\frac{\partial }{\partial t}\Delta \varphi +\left[ \left( -\mathbf{\nabla }%
\varphi \times \widehat{\mathbf{e}}_{z}\right) \cdot \mathbf{\nabla }\right]
\Delta \varphi =0 
\]

\item It is known that this equation is equivalent to a model of point-like
vortices interacting in plane by a self-generated long-range (Coulombian)
potential as known from prestigious theoretical work Kirchhoff, Onsager,
Montgomery, etc. (see \cite{kraichnanmontgomery}).

\item The dynamics of the point-like vortices has been mapped onto a field
theoretical model consisting of a complex scalar (matter) field $\phi $, $%
\phi ^{\dagger }$, and a gauge field $A^{\mu }$ (which mediates the
interaction) \cite{flmadisinh}. The Lagrangian density consists of the kinetic
term for $\phi $, the Chern-Simons term for $A^{\mu }$ and a nonlinear
interaction. The field variables are in the algebra $su\left( 2\right) $ due
to the vortical nature of the physical elements%
\[
\phi =\phi _{1}E_{+}+\phi _{2}E_{-} 
\]%
with $E_{\pm }$ the ladder generators in the cartan algebra. Introducing $%
\rho _{1}\equiv \left\vert \phi _{1}\right\vert ^{2}$, $\rho _{2}\equiv
\left\vert \phi _{2}\right\vert ^{2}$, the vorticity is%
\[
\omega =-\frac{2}{\kappa }\left( \rho _{1}-\rho _{2}\right) 
\]%
and in the relaxed states (self-dual) we have%
\[
\Delta \psi +\frac{4}{\kappa }\sinh \psi =0 
\]%
here $\psi $ is the streamfunction of the flow, with $\psi =\ln \rho $, $%
\rho _{1}=\rho _{2}^{-1}=\rho $.

\item We have noted that this equation is exactly the equation describing
the surfaces in real $3D$ space with the property of constant mean
curvature $H=$ const (CMC). If the curvatures of the surface are $\kappa _{1}$ and $\kappa
_{2}$ the CMC surfaces have%
\[
H=\frac{\kappa _{1}+\kappa _{2}}{2}=\frac{1}{2} 
\]

\item The fact that the self-duality states and the CMC surfaces are
governed by the same equation \emph{sinh}-Poisson has a deep significance.
Extending the parallel between these two system at states which are not at
self-duality or at CMC, (\emph{i.e.} the dynamical states preceeding the
asymptotic relaxed states) we have established the following mapping%
\begin{eqnarray*}
\rho _{1} &=&\frac{\left( \kappa _{1}+\kappa _{2}\right) ^{2}}{\kappa
_{1}-\kappa _{2}} \\
\rho _{2} &=&\kappa _{1}-\kappa _{2}
\end{eqnarray*}

\item It then results that the umbilic points of a surface in real $3D$
space, \emph{i.e.} a point where%
\[
\kappa _{1}=\kappa _{2} 
\]%
is a point where%
\begin{eqnarray*}
\rho _{1} &\rightarrow &\infty \\
\rho _{2} &\rightarrow &0
\end{eqnarray*}%
and this means that%
\[
\left\vert \omega \right\vert \rightarrow \infty 
\]%
at these points.

\item Now we see that looking for the points where there is a very high
value of the vorticity (almost singular vortices $\left\vert \omega
\right\vert \rightarrow \infty $) means to look for the umbilic points ($%
\kappa _{1}=\kappa _{2}$) of a surface.

\item We actually consider a turbulent region where the strong,
cuasi-singular vortices appear at random. Equivalently, we must consider a
surface in $3D$ which has fluctuating shape. The statistical
properties of the random strong vortical structures arising in turbulence is
the same as the statistical properties of the set of umbilic points on the
fluctuating surface. On the other hand the statistics of the fluctuations of the surface is assumed Gaussian.

\item We then have to find the statistical properties of the set of umbilic
points on a surface whose shape fluctuates with the Gaussian statistics.
\end{enumerate}

\bigskip

This problem is solved in Ref. \cite{berry}. For a Gaussian correlation of
the surface heigth $\rho \sim \exp \left( -r^{2}/\lambda ^{2}\right) $, the
density of umbilic points is%
\[
n=\frac{3}{\pi \lambda ^{2}} 
\]%
and this is also the density of high amplitude robust drift-type, vortices
generated by turbulence. Here $\lambda $ is the linear
dimension of a typical eddy of the drift wave turbulence, normalized to $%
\rho _{s}$.

\section{Appendix. Details about parameters of the exact solutions}
\renewcommand{\theequation}{A.\arabic{equation}} \setcounter{equation}{0}

The solution written in Eqs.(\ref{eq10}) and (\ref{eq11}) have been obtained
by Chow \cite{chow} using the Hirota method. The parameters are defined with
the expression of $A$ written in terms of the Riemann $\theta _{i}\left(
z,\tau \right) $-functions $i=1,2,3,4$, the general solution. The connection
with the original expression in terms of two "spatial" coordinates $\left(
x,y\right) $ consists of replacing, for our particular case%
\begin{eqnarray*}
k_{y} &\rightarrow &x \\
k_{x} &\rightarrow &y
\end{eqnarray*}%
Then, one has 
\[
\left\vert A\right\vert ^{2}=\lambda _{1}^{2}\left\vert \frac{\theta
_{4}\left( \alpha x,\tau \right) \theta _{1}\left( \beta y,\tau _{1}\right)
+\theta _{4}\left( \alpha x,\tau \right) \theta _{1}\left( \beta y,\tau
_{1}\right) }{\theta _{4}\left( \alpha x,\tau \right) \theta _{4}\left(
\beta y,\tau _{1}\right) +\theta _{1}\left( \alpha x,\tau \right) \theta
_{1}\left( \beta y,\tau _{1}\right) }\right\vert ^{2} 
\]%
The variables $\tau ,\tau _{1}$ are pure imaginary parameters of the \emph{%
nome} $q=\exp \left( i\pi \tau \right) $ which defines the expression of the
theta $\theta _{i}$ functions as series in powers of $q^{2}$. The relations
between the space scaling factors $\left( \alpha ,\beta \right) $ of the $x$
and $y$ variables and the two \emph{nome} variables is%
\[
\alpha \left[ \theta _{3}^{2}\left( 0,\tau \right) -\theta _{2}^{2}\left(
0,\tau \right) \right] =\beta \left[ \theta _{3}^{2}\left( 0,\tau
_{1}\right) -\theta _{2}^{2}\left( 0,\tau _{1}\right) \right] 
\]%
The coefficient is%
\[
\lambda _{1}^{2}=\beta ^{2}\theta _{2}^{2}\left( 0,\tau _{1}\right) \theta
_{3}^{2}\left( 0,\tau _{1}\right) -\alpha ^{2}\theta _{2}^{2}\left( 0,\tau
\right) \theta _{3}^{2}\left( 0,\tau \right) 
\]%
It is possible to replace the $\theta $-functions with Jacobi elliptic
functions and so new notations are introduced:%
\[
r=\alpha \theta _{3}^{2}\left( 0,\tau \right) \ \ ,\ \ s=\beta \theta
_{3}^{2}\left( 0,\tau _{1}\right) \ \ ,\ \ k=\theta _{2}^{2}\left( 0,\tau
\right) /\theta _{3}^{2}\left( 0,\tau \right) \ \ ,\ \ k_{1}=\theta
_{2}^{2}\left( 0,\tau _{1}\right) /\theta _{3}^{2}\left( 0,\tau _{1}\right) 
\]%
$K\left( k\right) \ $\ and $E\left( k\right) $ are the complete elliptic
integrals of modulus $k$. We see that the new (stretched) spatial variables
are $rx$ and $sy$. Since in this solution $x$ and $y$ are regarded as
spatial coordinates, their coefficients $r$ and $s$ are called
"wavenumbers". In our use of this solution the situation is reversed, $%
x\rightarrow k_{y}$ and $y\rightarrow k_{x}$ which means that $r,s$ are
spatial wavlengths along these directions.

The restriction exists: $k_{1}>k$ and the \emph{long wavelength} limit is 
\[
k_{1}\rightarrow 1 
\]%
leading to%
\[
s=r\left( 1+k\right) /2 
\]

For definitions and properties of the Riemann $\theta _{i}$ functions see 
\emph{Abramovitz and Stegun} \cite{AS} or \emph{NIST Digital Library of
Mathematical Functions} \cite{DLMF}.

\end{appendices}

\bigskip

\newpage

\bigskip

\begin{figure}[tbph]
\centerline{\includegraphics[height=10cm]{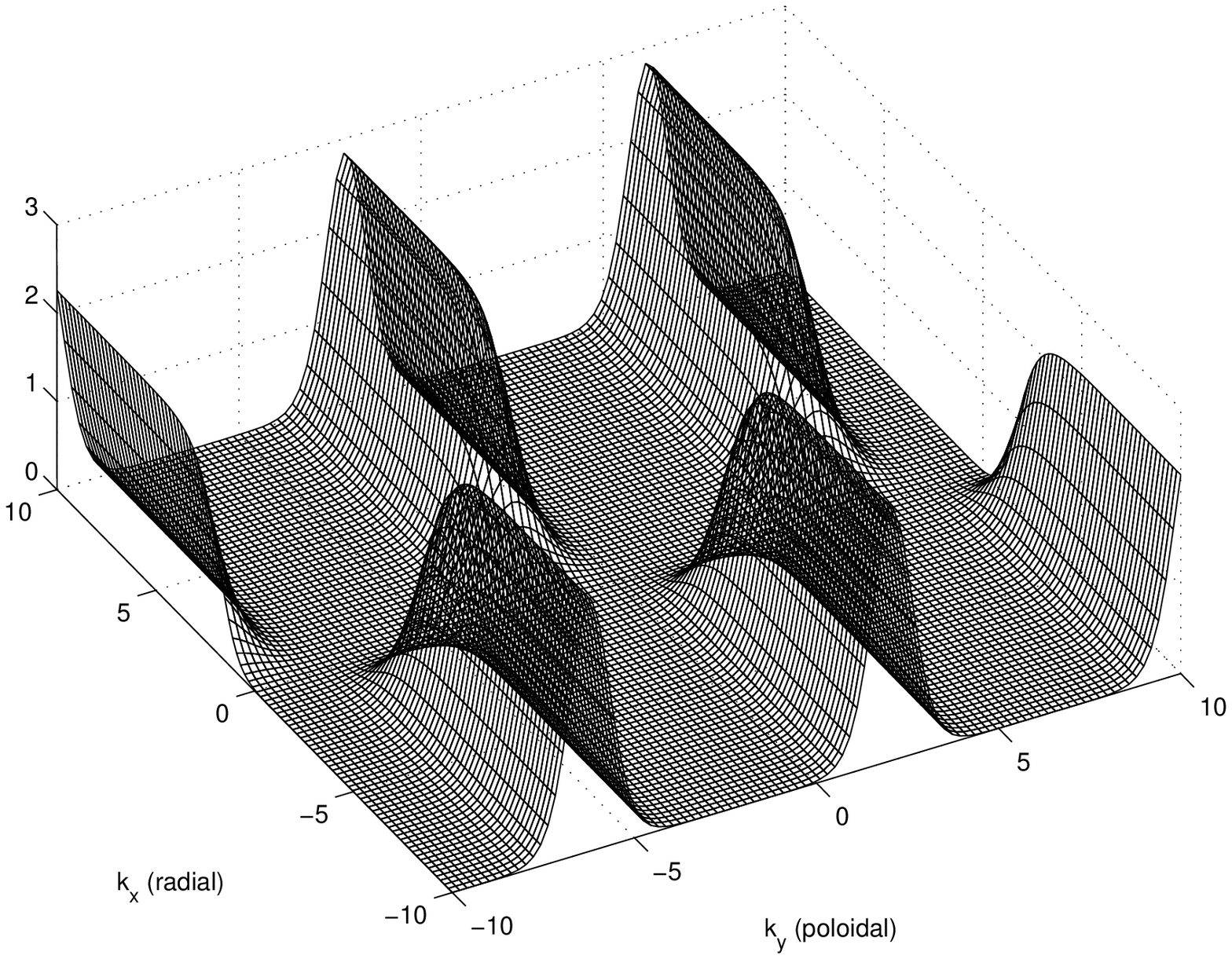}}
\caption{Graphical representation of the solution of the Devay Stewartson
system for the parameters $k=0.7$ and $r=1$.}
\label{figA2}
\end{figure}

\bigskip


\bigskip

\bigskip

\begin{thebibliography}{99}
\bibitem{flmadi1} F. Spineanu and M. Vlad, Nucl. Fusion \textbf{52} (2012)
114019.

\bibitem{xu} M. Xu, R.G. Tynan, P.H. Diamond \emph{et al.} Phys. Rev.Lett.%
\emph{\ }\textbf{107} (2011) 055003.

\bibitem{flmadiphenom} F. Spineanu, M. Vlad, K. Itoh, S.-I Itoh, \emph{%
http://arxiv.org/pdf/physics/0311138}.

\bibitem{drake} J.F. Drake, J.M. Finn, P. Guzdar \emph{et al.} Phys. Fluids 
\textbf{B4} (1992) 488.

\bibitem{shapiro1} R.Z. Sagdeev, V.D. Shapiro, V.I. Shevchenko,
Sov.J.Plasma.Phys. \textbf{4(3) }(1978) 306.

\bibitem{shapiro2} G.I. Soloviev, V.D. Shapiro, R.C.V. Somerville, B.
Shkoller, J. Atmos. Sci. \textbf{53} (1996) 2671.

\bibitem{DritschellScott} R.K. Scott and D.G. Dritschell, J. Fluid Mech. 
\textbf{711} (2012) 576.

\bibitem{BouchetSimonnet} F. Bouchet and E. Simonnet, Phys. Rev. Lett. 
\textbf{102} (2009) 094501.

\bibitem{ZakharovLvovFalkovich} V.E. Zakharov, V.S. L'vov, G. Falkovich, 
\emph{Kolmogorov Spectra of Turbulence I}, Spinger Verlag, Berlin, 1997.

\bibitem{HohenbergHalperin} P.C. Hohenberg, B.I. Halperin, Rev. Mod. Phys. 
\textbf{49} 91977) 435.

\bibitem{schecterdubin} D.A. Schecter and D.H.E. Dubin, Phys.Fluids \textbf{%
13} (2001) 1704.

\bibitem{montgomerysinh} D. Montgomery, W.H. Mathaeus, W.T. Stribling, D.
Martinez and S. Oughton, Phys. Fluids \textbf{A4}, (1992) 3.

\bibitem{flmadisinh} F. Spineanu, M. Vlad, Phys. Rev.E \textbf{67}, (2003)
046309.

\bibitem{KMcWT} R. Kinney, J. C. McWilliams and T. Tajima, Phys. Plasmas 
\textbf{2} (1995) 3623.

\bibitem{flmadiatm} F. Spineanu, M. Vlad, Phys. Rev. Lett. \textbf{94}
(2005) 025001 and \emph{http://arxiv.org/pdf/physics/0501020}.

\bibitem{wang} B. Wang, X. Li and L. Wu, Journal of Atmospheric Sciences 54
(1997) 1462.

\bibitem{marcus} P.S. Marcus, J. Fluid Mech. \textbf{215} (1990) 393.

\bibitem{linbarrancomarcus} H. Lin, J.A. Barranco, P.S. Marcus, Center for
Turbulence Research, Annual Research Briefs 2002, Stanford University.

\bibitem{schecterdubin2} D.A. Schecter, D.H.E. Dubin, Phys. Rev. Lett. 
\textbf{83 }(1999) 2191.

\bibitem{flmadi2} F. Spineanu and M. Vlad, Phys. Rev. Lett. \textbf{89}
(2002) 185001, and: \emph{http://arxiv.org/pdf/physics/0204050}.

\bibitem{chow} K.W. Chow, Wave Motions \textbf{35} (2002) 71.

\bibitem{kraichnanmontgomery} R.H. Kraichnan and D.C. Montgomery,
Rep.Prog.Phys.\textbf{43} (1980) 547.

\bibitem{berry} M.V. Berry and J.H. Hannay, J. Phys.A: Math. Gen., \textbf{10%
} (1977) 1809.

\bibitem{AS} M. Abramowitz and I. A. Stegun, Handbook of Mathematical
Functions, National Bureau of Standards Applied Mathematics Series 55,
(Tenth printing) 1972, Washington D.C.

\bibitem{DLMF} NIST Digital Library of Mathematical functions, \emph{%
http://dlmf.nist.gov}.
\end{thebibliography}
\end{document}